%% file: main-page.tex
\documentclass[sigconf]{acmart}
\usepackage{lipsum}
\usepackage{graphicx}
\usepackage{multirow}
\usepackage{adjustbox}
\AtBeginDocument{%
  \providecommand\BibTeX{{%
    \normalfont B\kern-0.5em{\scshape i\kern-0.25em b}\kern-0.8em\TeX}}}

\usepackage{array}
\newcolumntype{L}{>{\centering\arraybackslash}m{2cm}}
\newcolumntype{l}{>{\centering\arraybackslash}m{1.5cm}}
\newcolumntype{M}{>{\centering\arraybackslash}m{2.5cm}}
\newcolumntype{N}{>{\centering\arraybackslash}m{0.5cm}}
\setcopyright{acmcopyright}
\copyrightyear{2021}
\acmYear{2021}
\acmDOI{xx.xxxx/xxxxxxx.xxxxxxx}

\acmConference[DSSG '21]{The 3rd KDD Workshop on Data Science for Social Good}
\acmBooktitle{The 3rd KDD Workshop on Data Science for Social Good}

\begin{document}

\title{Exploring the Scope of Using News Articles to Understand Development Patterns of Districts in India}

\author{Mehak Gupta}
\email{mehak.gupta.0503@gmail.com}
\affiliation{%
 \institution{Indian Institute of Technology}
 \city{Delhi}
 \state{India}
}

\author{Shayan Saifi}
\email{shayansaifi@gmail.com}
\affiliation{%
 \institution{Indian Institute of Technology}
 \city{Delhi}
 \state{India}
}

\author{Konark Verma}
\email{mcs182025@iitd.ac.in}
\affiliation{%
 \institution{Indian Institute of Technology}
 \city{Delhi}
 \state{India}
}

\author{Kumari Rekha}
\email{mcs182144@iitd.ac.in}
\affiliation{%
 \institution{Indian Institute of Technology}
 \city{Delhi}
 \state{India}
}
\author{Aaditeshwar Seth}
\email{aseth@cse.iitd.ac.in}
\affiliation{%
 \institution{Indian Institute of Technology}
 \city{Delhi}
 \state{India}
}

\begin{abstract}
Understanding what factors bring about socio-economic development may often suffer from the \emph{streetlight effect}, of analyzing the effect of only those variables that have been measured and are therefore available for analysis. How do we check whether all worthwhile variables have been instrumented and considered when building an econometric development model? We attempt to address this question by building unsupervised learning methods to identify and rank news articles about diverse events occurring in different districts of India, that can provide insights about what may have transpired in the districts. This can help determine whether variables related to these events are indeed available or not to model the development of these districts. We also describe several other applications that emerge from this approach, such as to use news articles to understand why pairs of districts that may have had similar socio-economic indicators approximately ten years back ended up at different levels of development currently, and another application that generates a newsfeed of unusual news articles that do not conform to news articles about typical districts with a similar socio-economic profile. These applications outline the need for qualitative data to augment models based on quantitative data, and are meant to open up research on new ways to mine information from unstructured qualitative data to understand development. 
\end{abstract}


\maketitle

\section{Introduction}
\input{introduction}

\section{Related work}
\input{related-studies}

\section{Dataset}
\input{dataset.tex}

\section{Methodology}
\input{methodology.tex}

\section{Evaluation through contrasting patterns}
\input{evaluation.tex}

\section{Applications}
\input{applications.tex}

\input{newsfeed}

\section{Conclusions}
\input{conclusion.tex}

\bibliographystyle{ACM-Reference-Format}
\bibliography{references.bib}

\section*{Appendices}
\input{appendix}

\end{document}

%% file: introduction.tex
Indian districts demonstrate a wide diversity in patterns of socio-economic development and provide a natural experimental setting to understand the relationships between different development indicators \cite{district-development-model}. The discipline of development economics has traditionally used survey and census methods to measure such indicators, which are then fed into econometric models to find causal and correlated relationships between them. Doing manual surveys at regular intervals however is not only expensive, the survey methods and indicators can also change over time making it hard to obtain comparable data \cite{gdp-series-problems}. Several modern approaches using big-data sources such as satellite data, social media, and mass media, are now used to build proxy indicators for development to counter some of these problems \cite{nightlights, chahat-comad, economic-uncertainly, labour-market-flows}. All these methods however require a prior identification of development variables that are collected or obtained through proxies, and then analyzed for their inter-relationships with one another. Development is a complex process though and may be affected by variables that were not originally measured or considered worthwhile as influencing development or getting affected by it \cite{rodrik}. For example, the effect that communal violence in a location may have on development, say by affecting the inclination of migrant workers to relocate there for employment, or the effect that industrialization may have on environmental factors such as air or water pollution, may go unobserved if a prior hypothesis is not framed to measure these variables and test for their influence. 

The research question we attempt to investigate is whether qualitative data such as through news articles or social media that report on various events in different locations, can provide helpful hints of how these events may affect different socio-economic development patterns. This can in turn provide useful inputs to researchers for new hypotheses to frame or additional quantitative data to collect, to measure indicators based on these events and find out their relationship with development. Or at the very least, if not identify new information then such mass media and social media sources can provide qualitative data to portray the lived experiences of people experiencing different kinds of development patterns. For example, are districts that experience rapid industrialization also marked by news about worker protests, or are less developed districts marked by news about corruption in the delivery of welfare benefits? The wider motivation for our work arises from the need to look beyond well recognized quantitative indicators of development, to understand what these indicators mean in practice for the people \cite{thick-data}, and what other factors may have been overlooked in the conceptualization of development \cite{development-as-freedom}. 

We use a unique dataset to study this research question. In prior work, we built a machine-learning based classifier that uses satellite data from the Landsat-8 system as a proxy to predict census-based development indicators such as asset ownership of households, construction material used in residential housing, access to drinking water, etc \cite{chahat-comad}. The classifier was trained using data from the Indian census for 2001 and 2011, and used to obtain development indicators as of 2019, ahead of the next planned Indian census in 2021 \footnote{Once a census is conducted, it takes several more years for the government to collate and release the data.}. These indicators were linearly summed to build a development index we termed as ADI (Aggregate Development Index), similar to the HDI (Human Development Index) methodology that combines economic, literacy, and health based indicators into a common measure for broad-based development \cite{hdi}. In addition, the Indian census of 2011 served to categorize the districts on their primary source of employment, as non-agricultural or agricultural or with high unemployment. We use these two variables of the type of employment, and change in ADI between 2011 and 2019, to obtain different categories of Indian districts that underwent different development patterns, such as industrial districts that experienced a large change in ADI over the last decade, or agricultural districts that did not, and so on. Our methodology is generic though and can be applied to other categorizations as well, for example, to categorize districts based on health or income or education variables, and then understand the pace of growth in these categories. In parallel, we have also built a large corpus of news articles from six primary English language newspapers in India, since 2011 until now, consisting of all articles published on a daily basis \cite{ictd-media, compass-media}. The district location of these articles is obtained through an entity extraction method, and the media corpus thus serves as the source of qualitative data to understand the different development patterns. 

Investigating our research question essentially amounts to solving a document ranking problem, to rank news articles about districts that are most representative of the development pattern of that district, and most distinct from articles about districts that underwent other development patterns. We evaluate different ranking methods based on TFIDF (Term Frequency Inverse Document Frequency), LDA (Latent Dirichlet Allocation, a topic-modeling method), and document vectors (an auto-encoder based deep learning approach). We use these methods to demonstrate several applications, such as to identify news articles that best explain the development of districts which are outliers from the dominant development pattern to which they belong, contrast pairs of districts that were at a similar development status in 2011 but have diverged significantly since then, and a newsfeed application to identify interesting news articles that may be counter-intuitive to the dominant pattern of development of their district. Our analysis shows that news articles are a promising source of qualitative data to understand the development of districts in India, and highlights the scope that bottom-up qualitative data has to complement patterns uncovered by top-down big data analysis.

%% file: related-studies.tex
A highly public debate has raged since several years about what constitutes socio-economic development \cite{bhagwati-sen}. On the one hand, economic growth is seen as the ``tide that lifts all boats'' and needs to be a precursor before equitable benefits of health, nutrition, education, dignified livelihood, etc. can reach people \cite{bhagwati-book}. On the other hand, instrumental freedoms such as good health, education, and food security are seen as essential elements that need to be serviced through public expenditure, to achieve constitutive freedoms for equitable economic development \cite{development-as-freedom}. The diversity of social and economic development across different Indian districts can potentially serve as a natural experiment to inform this debate, but comprehensive datasets to build different socio-economic indicators over time do not exist \cite{district-development-model}. Attempts have been made to use proxy indicators such as nightlights \cite{nightlights}, but ultimately any econometric model will be as good as how comprehensive are the factors that it models. Our goal in this paper is to inform what factors relevant to development may be missing in development models, or can at least provide a qualitative understanding of what these indicators mean in terms of the lived experiences of people. 



Our enthusiasm about the importance of qualitative data to supplement quantitative data is shared by people like Tricia Wang who highlight the importance of ethnography in generating ``thick data'' to bridge the social context-loss between data points obtained through big-data analysis \cite{thick-data}. Qualitative datasets such as through social media, mass media, or Wikipedia pages, can play a role in revealing the underlying dynamics that cannot be directly observed in large quantitative datasets. Lu et al. similarly identify correlations between green buildings and construction waste management using a triangulation of quantitative big data obtained from government agencies with qualitative ``thick data'' obtained from case studies and interviews \cite{article_80}. They highlight the importance of understanding the underlying context before blindly promoting any system such as the Green Building Rating System (GBRS). Similarly, a report from IBM writes about the strategy of ``mixed analytics'' which combines ethnographic practices with big data analysis in the domain of public services \cite{article_9}. In recent years, social media platforms have also been considered as a key source to record social interactions rather than the traditional ways of ethnography in the form of surveys, group discussions, and interviews. People tend to be vocal about their experiences and opinions on social media and can facilitate new approaches of digital ethnography \cite{article_26, article_19}. 

These research directions of using qualitative data to explain or augment quantitative data is different from some other recent work in improving quantitative predictions using qualitative data. Researchers have used Wikipedia pages, for example, to improve satellite-data based predictions of development \cite{wikipedia-satellite}. Similarly, news articles about weather events have been used to improve the predictions of agricultural commodity prices \cite{sunandan-agri-commodities}, and methods have been developed for event identification from news articles \cite{sunandan-socio-economic-events}. News articles have also been used to build indicators for economic uncertainty \cite{economic-uncertainly}, and similarly social media data has been used to build indicators of unemployment \cite{unemployment}, and Google search trends have been used to predict flu outbreaks \cite{google-flu}. Our work is distinct and complementary to this research direction. We use similar techniques for text analytics, such as topic modeling using LDA \cite{topic-modeling}, or document vectors using deep learning approaches \cite{doc2vec, docvec}, but we do not aim to solve a prediction problem. Our work is closer to a ranking problem to identify news articles that are representative of specific development patterns, and at the same time also distinct from news articles about other development patterns. These news articles can then give hints about more extensive surveys or data collection exercises that should be commissioned to enrich development models with new relevant variables.

%% file: dataset.tex
We next describe our dataset that comprises of three parts. First, to obtain a categorical variable as of 2011 of the primary type of employment in a district: agricultural, non-agricultural, or high unemployment. Second, to obtain a categorical variable of the pace of socio-economic development of a district from 2011 to 2019: fast, medium, slow, based on the relative change in ADI (Aggregate Development Index). Third, to build a district-wise corpus of news articles published between 2011 to 2019. 


\subsection{Employment type}

The Government of India conducts a household-level population census every 10 years. We used data from the 2011 census, available from its official website \cite{census}, to build a categorical variable of the primary type of employment in each district. The census reports for each district the population employed in agriculture, construction, mining, technical services, public administration, and other industry sectors. We first grouped these mutually exclusive variables into three broad parameter types of unemployment (unemp), agricultural (agri), and non-agricultural (non-agri) employment. For example, cultivators and agricultural workers were considered as employed in agriculture, other types of services and trades were grouped together as non-agricultural employment, and the rest of the adult non-working population was considered as not employed. We then did a k-means clustering of the districts based on the vector of the percentage of population in these three categories: agricultural, non-agricultural, and unemployed. Figure \ref{fig:boxplot} shows a box-plot for the distribution of districts across these categories (k = 3). The unemp category contains districts where there is heavy unemployment and significantly less agricultural or non-agricultural employment. The rest of the districts are split into agri and non-agri categories depending upon which one of them is more dominant. A detailed explanation of the discretization method and robustness for different values of k can be found in \cite{district-development-model}. 

\begin{figure}
    \centering
    \includegraphics[width=0.8\columnwidth, height=5cm,scale=0.7]{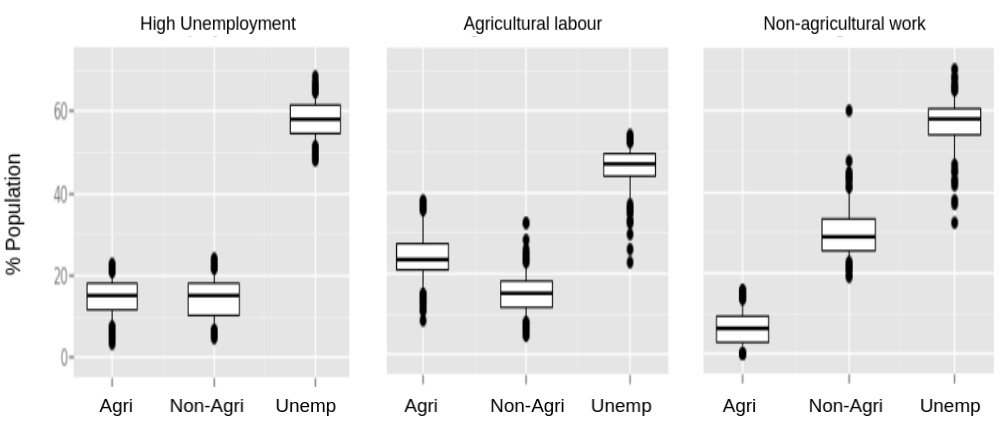}
    \caption{\centering Distribution of the \% of population under different types of employment (unemployed, agricultural, non-agricultural) across three district clusters}
    \label{fig:boxplot}
\end{figure}

\subsection{Pace of growth}

Similar to the district-wise population distribution of employment in different industry sectors, the census reports seven other indicators: literacy, access to drinking water, access to sanitation facilities, fuel used for cooking, main source of lighting, asset ownership, and construction condition of households. Each of these indicators is reported in terms of constituent variables, for example, the fuel used for cooking is reported with the number of households that use firewood, kerosene, LPG (Liquid Petroleum Gas), PNG (Piped Natural Gas), and biogas. We group these variables into three broad parameter types of rudimentary (RUD), intermediate (INT), and advanced (ADV) forms of fuel used for cooking. Firewood is considered as a rudimentary type of fuel, kerosene and cow dung are grouped together as an intermediate type, and PNG, LPG and bio-gas are grouped as advanced types of fuels for cooking. Following a method similar to the one used by Fabini et al. \cite{kenya-electricity-use} to discretize a collection of constituent variables, a k-means clustering is done on this 3-tuple vector (\% RUD, \% INT, \% ADV) to obtain an ordinal label for each districts as level-1/2/3 in terms of the primary type of fuel for cooking used in the district. Figure \ref{fig:boxplot} shows a box-plot for the districts mapped to these three levels (k = 3). Level-1 districts predominantly use rudimentary types of fuel for cooking, level-2 districts use intermediate types of fuel, and level-3 districts use advanced types of fuel for cooking. A detailed explanation of the discretization method and its robustness for different values of k is given in prior work \cite{district-development-model}. In a similar manner, each of the seven indicators are used to label districts in terms of three levels. 


\begin{figure}
    \centering
    \includegraphics[width=0.8\columnwidth, height=5cm,scale=0.3]{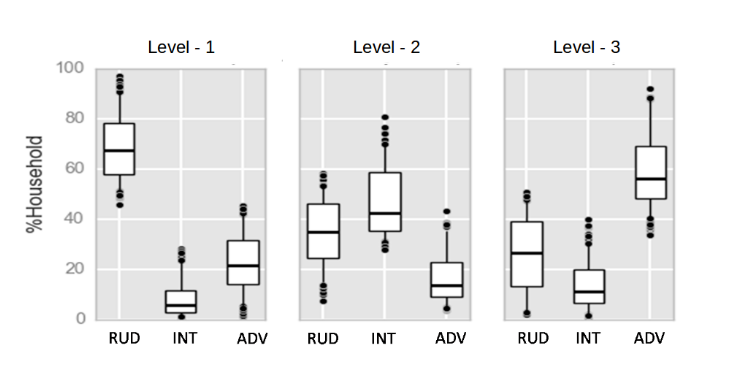}
    \caption{\centering An example of discretizing the level of development of districts in terms of the fuel used for cooking in the district households. Shown is the distribution of the \% of households using different types of fuel rudimentary (RUD), intermediate(INT), or advanced(ADV) across three district clusters.}
    \label{fig:boxplot}
\end{figure}

In another prior work, we used these district labels obtained from the Indian census of 2001 and 2011 to build a machine learning classifier using satellite data \cite{chahat-comad}. This classifier uses daytime satellite imagery obtained from the Landsat-8 system (open dataset available through the Google Earth Engine), to obtain the corresponding socio-economic development level for the district under the seven different indicators. We used this classifier to take the satellite imagery as of 2019 as an input, and obtain a prediction of the current levels of socio-economic development of all the Indian districts since a census has not been conducted in recent years as yet. An evaluation of the model on the ten-yearly data from 2001-2011 gave a steady performance of between 70-80\% accuracy for each of the seven indicators. Interestingly, we found that a simple summation of the levels of all these indicators to create an Aggregate Development Index (ADI), on the same lines as how the HDI (Human Development Index) is computed as a summation of three indicators on life expectancy, literacy rate, and per capita GDP, gave a low normalized RMSE of 0.04, likely due to a cancellation of errors made in the prediction of different indicators. We were thus able to deploy the classifier to obtain the ADI as of 2019, as a value between 7 (all seven indicators at level-1) to 21 (all seven indicators at level-3).

\begin{figure}
    \centering
    \includegraphics[scale = 0.4]{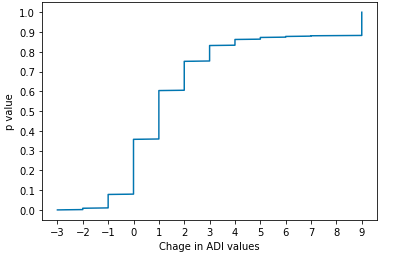}
    \caption{CDF Plot for change in ADI values}
    \label{fig:cdf}
\end{figure}

We then use the difference in ADI between the predicted values for 2019 and the known values for 2011 to calculate the pace of socio-economic development for each district. The difference in ADI values was in the range -3 to 9, as shown in the CDF plot in Figure \ref{fig:cdf}. We consider the bottom 36\% of the districts as slow developing districts (with a change in ADI between -3 to 0), the next 37\% as medium developing (with a change in ADI between 1 to 3), and last 27\% as fast developing districts (with a change in ADI between 4 to 9). In addition, those districts that had a very high ADI value of 20 or 21 for both 2011 and 2019, were considered as fast growing districts as well. 
Thus, by categorizing districts on the two axes of primary type of employment (agri, non-agri, unemp) and pace of growth (fast, medium, slow) we are able to obtain 9 sub-classes of districts that underwent different development patterns, as shown in Table \ref{tab:subclass}. 

\begin{table}[H]
    \centering
    \begin{tabular}{|c|c|c|c|}
    \hline
          & Slow & Average & Fast  \\
    \hline
    Unemp &  97 (92) & 100 (97) & 31 (31)   \\
    \hline
    Agri  &  91 (90) & 102 (101) & 34 (34)   \\
    \hline
    N-Agri &  23 (23) & 30 (30) & 81 (81)   \\
    \hline
    \end{tabular}
    \caption{Total number of districts in each subclass (number of districts with enough news articles)}
    \label{tab:subclass}
\end{table}
\raggedbottom


\subsection{Mass media dataset}
We next build a corpus of news articles for each of the 9 sub-classes. Through web-crawling of archives from January 1st 2011 and ongoing daily data collection, we have built a large corpus of news articles containing all the news published by six prominent national English newspapers in India: Hindustan Times, Times of India, The Hindu, The Indian Express, Deccan Herald, and the New Indian Express. This corpus contains more than 5 million articles. In prior work, we used this corpus to study media bias in the coverage of several important economic and technology policies in India \cite{ankur-mass-media, compass-media, ictd-media}. 

For this paper, we use the corpus to isolate news articles along the following five broad themes: agriculture, industrialization, social welfare, environment, and politics and social activities. We chose these themes because most analysis of development revolves around them \cite{uncertainglory}. Agriculture is considered as the primary industrial sector of the Indian economy. Manufacturing and services are the secondary and tertiary sectors, which we clubbed together under the industrialization theme. Social welfare is a strong determinant of development, comprising almost 10\% of the Indian GDP. Environment is a cross-cutting theme which has become an increasingly important category and mediates many development strategies. The final theme of social activities and politics was selected as reflective of what aspects of development are expressed and considered important by society. This includes topics like crime, traffic, communal violence, sports, and elections. 

We developed a simple keyword based method to map news articles to one or more of these themes. An initial seed set of keywords was put together by the authors based on their knowledge of the themes, and a query expansion process was then followed to identify new keywords from among frequently occurring keywords in the current document selections. New documents corresponding to the expanded queries were then obtained, with up to three iterations after which saturation was noticed for all the themes. The final document collections were made for the five themes by selecting news articles that contained at least three keywords from the keyword set for that theme. The keywords used are given in Table \ref{tab:keywords} in the Appendix. Articles with overlapping themes were mapped to multiple themes. 

Once the theme-wise collections were obtained, entity extraction of the district location was used to identify the districts referred to in a news article. We first used the OpenCalais service \cite{opencalais} to identify locations mentioned in the article, and then followed several steps to resolve these location entities against the list of districts as per the Indian census. We only considered entities mentioned in the opening paragraph of an article, but excluded the news bureau location which is typically prefixed at the start of an article. We then ran an edit distance match between the extracted location entities and the census list of districts. Any entities with a poor match were manually examined to check for spelling variations. We also took care of cases where districts could be referred to by multiple names (such as \emph{Poorvi}/\emph{Pashim} Champaran for \emph{East}/\emph{West} Champaran respectively), or the names of districts were changed (such as \emph{Prayagraj} for \emph{Allahabad}), or a district was split into smaller districts (such as \emph{Thane} was split into \emph{Thane} and \emph{Palghar}). In the case of split districts, we only considered the merged entities as of the 2011 census. Through an iterative process, we thus built a comprehensive district resolution list, and upon evaluating 100 randomly chosen entities we obtained precision and recall values of 93.18\% and 95.35\% respectively \cite{entity-resolution}. 

The district mapping thus enabled us to categorize the theme-wise document collections into the 9 sub-classes based on the type of employment and pace of growth seen in the districts. Table \ref{tab:articles} shows the count of news articles for the five themes against the 9 sub-classes. Finally, some pre-processing steps were undertaken on the documents for stop-word removal, stemming, and entity blinding so that the subsequent models we build using the documents are agnostic to names of people, organizations, political parties, etc. We also did not consider those districts for which the number of news articles was less than 100. 


\begin{table*}[t]
    \centering
    \small
    \begin{tabular}{|c|c|c|c|c|c|c|c|c|c|c|}
    \hline
        \multirow{2}{*}{Collections} & 
        \multirow{2}{*}{\begin{minipage}{0.5in}\centering Total articles \end{minipage}} & \multicolumn{3}{c|}{Unemp districts} & \multicolumn{3}{c|}{Agri districts} & \multicolumn{3}{c|}{Non Agri districts}\\
        \cline{3-11}
                    &                    & Slow & Medium & Fast & Slow & Medium & Fast & Slow & Medium & Fast\\ 
        \hline
        Agriculture & 128487 & 11097  &  5968  &  3968  &  9639  &  12204  &  4183  &  6762  &  5085  &  69581\\
        \hline
        Social welfare & 25709 & 2160  &  1085  &  647  &  1373  &  1565  &  570  &  1633  &  1199  &  15477 \\
        \hline
        Environment & 155205 & 10248  &  6662  &  5020  &  7734  &  8665  &  3518  &  9776  &  5890  &  97692\\
        \hline
        Industrialization & 172430 & 8379  &  4568  &  3648  &  5798  &  6269  &  2084  &  8337  &  4964  &  128383\\
        \hline
        Politics and social activities & 370509  & 29770  &  13957  &  10458  &  14513  &  12628  &  5238  &  26743  &  12748  &  244454\\    
        \hline
    \end{tabular}
    \caption{Total number of articles in each subclass}
    \label{tab:articles}
\end{table*}
\raggedbottom

%% file: methodology.tex
\label{ch:methodology}

Based on district-wise categorical variables of the primary type of employment and the pace of growth of a district, we define three types of development patterns that we attempt to investigate: 
\begin{itemize}
    \item \emph{Pattern 1:} To understand what factors relate to a different pace of growth among high-unemployment districts
    \item \emph{Pattern 2:} Similarly, among agricultural districts
    \item \emph{Pattern 3:} And among non-agricultural districts 
\end{itemize}

Note that the framework we have outlined is flexible to define other development patterns as well based on which variables are to be contrasted with one another. Our objective is simply to evaluate through a demonstration of whether this approach of using mass media can help explain or provide qualitative insights about the dynamics of different development trajectories followed by the districts. For each pattern, we therefore solve a document ranking problem to identify news articles that are most representative of the sub-class to which they belong, and most different from other sub-classes in the pattern under investigation. We compare three methods for this purpose. 



\subsection{TFIDF based selection}
TFIDF is a common information retrieval technique to obtain a weighting factor for each word in a document in the corpus. Term frequency TF($w$, $d$) is the frequency of word $w$ in document $d$. IDF($w$, $D$) is the logarithmically scaled inverse of the number of documents in the corpus $D$ that contain word $w$. TFIDF($t$, $d$, $D$) is then computed as TF($w$, $d$) $*$ IDF($w$, $D$), and essentially gives a higher weightage to words that occur more frequently in a document, but also less frequently in the overall corpus so as to identify words that are unique to the given document. In our case, we concatenated all the articles of a sub-class to form one document for that sub-class, and then computed TFIDF scores for all the words in each sub-class document. From each sub-class, we then obtained top-100 news articles by summing the TFIDF scores of all the words in the news article. We then compared the quality of this selection of the top-100 articles with selections obtained using other competing methods, as described next.

\subsection{Topic modeling based selection}
LDA (Latent Dirichlet Allocation) is a common topic modeling technique that identifies hidden topics as having been generated through a mixtures of keywords, and a mixture of the topics as further generating the documents in the corpus. We ran LDA over the news articles for each sub-class to identify the constituent topics for each sub-class. Based on the probability for an article to belong to a topic as given by the LDA routine, we then picked up $100 / k$ articles from each topic, where $k$ was the optimal number of topics identified through LDA in the sub-class. If a topic has less than $100 / k$ articles, additional articles were then picked up equally from other topics to finally select the top-100 articles for a sub-class. We used topic coherency scores to determine a good value of $k$ \cite{coherency-score}. 

Our intuition behind sampling articles from across the different topics identified by LDA was to choose a diverse set of articles from the sub-class, so that diverse descriptions of common development processes for that sub-class could be obtained. As compared to the previously described TFIDF method, the LDA method is therefore likely to give a more complete representation of articles for a sub-class. However, this method does not select articles that are distinct from other sub-classes, which the IDF component of TFIDF is able to achieve. In the next method using document vectors, we try to achieve both goals of choosing diverse articles in a sub-class, and also picking up articles that are distinctive to the sub-class as compared to other sub-classes. 


\subsection{Clustering over vector embeddings}
This method works through a three-step process as follows.

\begin{enumerate}

\item \textit{Vector embeddings using DocTag2Vec}. 
Doc2Vec is a deep-learning based unsupervised approach to obtain a vector representation for a document \cite{doc2vec}. It is known to work better for information retrieval tasks than TFIDF based vector representations since it also takes the order of words into account, thus making it more sensitive to the context in which different words are used. An extension to Doc2Vec is the DocTag2Vec architecture which additionally allows modeling of document tags: The vector representation of documents thus becomes sensitive to the tags attached to the documents, and vectors are also obtained for the tags as well \cite{docvec}. In our case, we consider the categorical variables of the type of employment and the pace of growth as tags. Districts are also uniquely identified through tags. 


\begin{figure}[t]
    \centering
    \includegraphics[scale = 0.3]{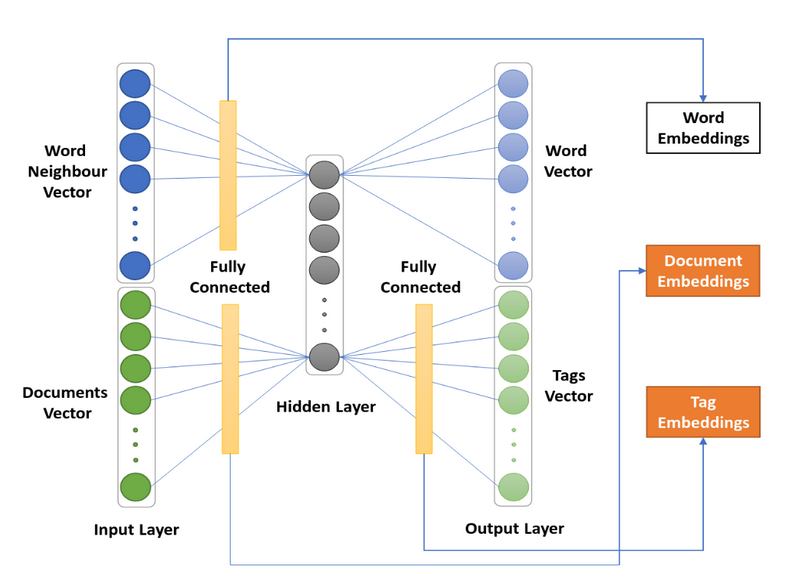}
    \caption{Doc-Tag-2-Vec Model \cite{docvec}}
    \label{fig:doctag2vec}
\end{figure}

Figure \ref{fig:doctag2vec} shows the DocTag2Vec architecture. Documents (news articles) and tags (type of employment, pace of growth, district) are taken as inputs, and corresponding embeddings are generated for the documents and tags. We trained this model separately for the article collections for each theme, thus obtaining five different models. The training parameters are described in more detail in the Appendix \textcolor{red}{\ref{ch:appendix}}. After empirical evaluation, 50-dimensional vectors were found to work well. 




\item \textit{Hierarchical clustering of document embeddings}. 
Within each sub-class, we use agglomerative hierarchical clustering on the DocTag2Vec document embeddings to obtain diverse topic clusters \cite{clustering-method}. We remove outlier clusters that contain very few articles, and finally obtain $k$ clusters for each sub-class where the value of $k$ is decided based upon a cophenetic distance statistic that determines which clusters can be merged with one another. Different sub-classes may have a different number of topic clusters.

\item \textit{Document selection}. Finally, we then compute a global centroid for each sub-class based on the document vectors of the news articles in the sub-class. Similarly, we compute a cluster centroid for the $k$ topic clusters in each sub-class. Based upon the appropriate ranking criteria for different applications, as we explain next, we are then able to use the document vectors, cluster centroid vectors, and sub-class centroid vectors, to identify the highest ranked news articles for the application. We calculate a score $s_{dt}$ as the cosine similarity between a document vector $d$ and the centroid of the topic cluster $t$ to which the document belongs, $s_{ds}$ as the cosine similarity between a document vector $d$ and the centroid of the sub-class $s$ to which the document belongs, and $s_{ds'}$ as the cosine similarity between a document vector $d$ and centroid of a contrasting sub-class $s'$. In applications where our goal is to select news articles that are most representative of their sub-class but most distinct from a contrasting sub-class, we rank the news articles based on the metric $s_{ds} * s_{dt} - s_{ds'}$ and select $100/k$ articles from each topic cluster in the sub-class $s$. Similarly, in applications where news articles need to be selected based on how much a district conforms to typical districts in its sub-class, we use the metric $s_{ds} * s_{dt}$ to rank news articles about that district. A document vector representation can thus be flexibly used for different applications. 


\end{enumerate}

%% file: evaluation.tex

We next compare the three methods of TFIDF, LDA, and DocTag2Vec, to identify the most suitable method for article selection and ranking for our primary use-case to identify articles most representative of their sub-class as well as most distinctive from other sub-classes. We evaluate using both quantitative metrics as well a qualitative user evaluation. 


For a quantitative evaluation, we use the intuition that the top-100 articles identified for a sub-class should have the least overlap of keywords with the top-100 articles from other sub-classes. For each pair of sub-classes, we therefore compute a Jaccard similarity coefficient on keywords in the articles. A low Jaccard similarity would imply that the article selection method was able to identify articles containing keywords quite distinct between pairs of sub-classes. Another metric we develop based on the same intuition is an entropy based measure that assigns a probability value to a keyword based on the most likely sub-class to which it belongs. A low entropy value for a sub-class would imply that the set of keywords of dominant use within it, are distinct from dominant keywords in other classes. We build two additional metrics as well based on TFIDF vectors for the top-100 articles in each sub-class: A mean cosine based and a Euclidean distance based similarity score between the articles in pairs of sub-classes. Low cosine similarities or high Euclidean distances would denote greater separation between the classes. The detailed values of the metrics are given in the Appendix, and Table \ref{tab:met_eval} shows the best performing method based of each of these four metrics. Clearly, the DocTag2Vec method seems to work the best for most cases. 

We also conducted qualitative evaluations for each of the three patterns, based on a self-assessment of the identified articles, and another assessment by volunteers from among graduate students in our research group. Table \ref{tab: pat1}, \ref{tab: pat2}, \ref{tab: pat3} shows the primary topics for each (sub-class, theme) pair identified by the authors through a manual inspection of the top identified articles. Several expected topics emerge for various development patterns, such as articles about irregularities in job schemes funded by the government, and crimes including thefts and violence, are common among slow-growing and high-unemployment districts (belonging to pattern-1). Fast-growing and high-unemployment districts had more news articles about industrial projects, development research, and festivals. Pattern-2 related to agricultural districts had topics such as farmer suicides and food adulteration surfacing among slow-growing districts, while fast-growing districts had more articles related to government schemes and good agricultural harvests. Topics like illegal mining and electricity power crises came up among slow-growing non-agricultural districts, while economic revival and revenue generation by the state were reported among fast-growing non-agricultural districts. 

We also conducted a user evaluation by recruiting 11 volunteers from among non-authors of this paper. A training session was held to communicate the purpose of the study and each rater was asked to assess based on 10 randomly chosen articles (from among the top-100 articles identified by the three ranking methods described earlier) for each pattern about whether the articles were representing commonly known development pathways within the patterns. Three questions were put forth to the raters for each theme, to assess based on a scale of 0..5 whether the articles were (a) relevant to the sub-classes, (b) they depicted events or topics that could be associated with positive or negative development, (c) the article selections for different sub-classes within each pattern was diverse and denoted different aspects. A 2-sample t-test was then conducted for each question between the ratings given for the different methods. In most cases, ratings given to the DocTag2Vec method came out as better than the other methods. The detailed t-test results are shown in Table \ref{tab:t_test} in the appendix.

\begin{table*}[t]
    \centering
    \begin{tabular}{|c|c|c|c|c|c|L|}
    \hline
                    Pattern & Metric  & Agriculture & Social welfare & Environment & Industrialization & Politics and social activities\\
    \hline  
    \multirow{3}{*}{Slow vs fast unemp} & 
    \begin{minipage}{1.2in} \strut \centering TFIDF based jaccard similarity  \strut \end{minipage} & DocTag2Vec & LDA & DocTag2Vec & DocTag2Vec & DocTag2Vec \\
    \cline{2-7}
                                 & \begin{minipage}{1.2in} \strut \centering Global similarity \strut \end{minipage} & DocTag2Vec & TFIDF & DocTag2Vec & DocTag2Vec & DocTag2Vec\\
    \cline{2-7}
                    & \begin{minipage}{1.2in} \strut \centering Euclidean distance \strut \end{minipage}          &     DocTag2Vec & TFIDF & DocTag2Vec & DocTag2Vec & DocTag2Vec\\
    \cline{2-7}
                    & \begin{minipage}{1.2in} \strut \centering Entropy \strut \end{minipage}          &     TFIDF & DocTag2Vec & TFIDF & TFIDF & LDA\\
    \cline{2-7}
                    & \begin{minipage}{1.2in} \strut \centering Qualitative evaluation \strut \end{minipage}          &    DocTag2Vec & LDA & TFIDF & DocTag2Vec & DocTag2Vec\\

    \hline
    
    \multirow{3}{*}{Slow vs fast agri} & 
    \begin{minipage}{1.2in} \strut \centering TFIDF based jaccard similarity  \strut \end{minipage} & DocTag2Vec & TFIDF & DocTag2Vec & DocTag2Vec & DocTag2Vec \\
    \cline{2-7}
                                 & \begin{minipage}{1.2in} \strut \centering Global similarity \strut \end{minipage} & LDA & DocTag2Vec & TFIDF & DocTag2Vec & DocTag2Vec \\
    \cline{2-7}
                    & \begin{minipage}{1.2in} \strut \centering Euclidean distance \strut \end{minipage}          & TFIDF & DocTag2Vec & DocTag2Vec & DocTag2Vec & LDA\\
    \cline{2-7}
                    & \begin{minipage}{1.2in} \strut \centering Entropy \strut \end{minipage}          &     DocTag2Vec & & TFIDF LDA & DocTag2Vec & DocTag2Vec\\
    \cline{2-7}
                    & \begin{minipage}{1.2in} \strut \centering Qualitative evaluation \strut \end{minipage}          &     DocTag2Vec & LDA & DocTag2Vec & DocTag2Vec & TFIDF\\
    \hline
    \multirow{3}{*}{Slow vs fast non-agri} & 
    \begin{minipage}{1.2in} \strut \centering TFIDF based jaccard similarity  \strut \end{minipage} & LDA & DocTag2Vec & TFIDF & LDA & DocTag2Vec \\
    \cline{2-7}
                                 & \begin{minipage}{1.2in} \strut \centering Global similarity \strut \end{minipage} & TFIDF & TFIDF & DocTag2Vec & LDA & LDA\\
    \cline{2-7}
                    & \begin{minipage}{1.2in} \strut \centering Euclidean distance \strut \end{minipage}          & TFIDF & TFIDF & DocTag2Vec & LDA & TFIDF \\
    \cline{2-7}
                    & \begin{minipage}{1.2in} \strut \centering Entropy \strut \end{minipage}          &     DocTag2Vec & LDA & LDA & LDA & DocTag2Vec\\
    \cline{2-7}
                   & \begin{minipage}{1.2in} \strut \centering Qualitative evaluation \strut \end{minipage}          &    DocTag2Vec & LDA & LDA & DocTag2Vec & DocTag2Vec\\
    \hline
    \end{tabular}
    \caption{Quantitative evaluation results}
    \label{tab:met_eval}
\end{table*}
\raggedbottom

%% file: applications.tex

Given the superior performance of the DocTag2Vec method in most cases, we next use this method to demonstrate a number of applications we can build from the models. 


\subsection{Contrasting pairs of districts}
The development indicators for 2011 and 2019 can identify districts that were at the same development status (in terms of ADI) in 2011 and with the same type of primary employment, but achieved a markedly different development status by 2019. An interesting application can be to find pairs of such districts that followed divergent development trajectories, and understand through the news articles about the events that transpired in the districts during the intervening years of 2011 to 2019 that may help explain this difference. 

An example is the pair of districts of Thiruvarur and Thanjavur in the state of Tamil Nadu, both of which were high unemployment districts in 2011 with an ADI of 16, but while Thiruvarur's ADI in 2019 only improved marginally to 17, the ADI of Thanjavur improved to 20. Table \ref{tab:d_analysis} shows the primary topics in the top ranked news articles for the two districts, based on the ranking metric of $s_{ds} * s_{dt} - s_{ds'}$ as described in Section \ref{ch:methodology}, for each of the five themes. The articles from Thanjavur are progressive in nature and include news about job schemes, tourism, agricultural waivers, and scholarships, while Thiruvarur has articles on crop damage, protests by political parties (\emph{bandhs}), and cyclones. Such an analysis method can therefore help provide a description of different development pathways followed by the districts. 

\begin{table*}[]
    \centering
     \small
    \begin{tabular}{|M|M|M|M|M|M|}
         \hline
         & Agriculture & Social welfare & Environment & Industrialization & Politics and social activities\\
         \hline
            Thiruvarur &  
        \begin{itemize} 
        
            \item Crop damage 
            \item Paddy cultivation
            \item Farmer suicide due to indebtedness
         
        \end{itemize}
       & \begin{itemize}
            \item Poultry Farms
            \item Road blockade
            \item Job reservation
        \end{itemize} 
        & \begin{itemize}
            \item Crop training
            \item Large scale strike
            \item Viral disease
        \end{itemize} 
      & \begin{itemize}
            \item Methane issue
            \item Cyclone
            \item Data theft
        \end{itemize} 
      & \begin{itemize}
            \item Diseases
            \item Crimes
            \item Adulteration
        \end{itemize} 
       \\
        \hline
        Thanjavur &  
        \begin{itemize} 
            \item Good monsoon
            \item Loan waiver
            \item Water conservation
        \end{itemize}
       & \begin{itemize}
            \item Check dams
            \item Entrepreneur-ship
            \item Job scheme
        \end{itemize} 
        & \begin{itemize}
            \item Tourism promotion
            \item Agri Technologies
            \item Scholarship
        \end{itemize} 
        & \begin{itemize}
            \item Handloom expo
            \item Entrepre-neurs
            \item Festivals
        \end{itemize} 
        & \begin{itemize}
            \item Cultural programme
            \item Tourist activities
            \item Dengue control
        \end{itemize} 
        \\
        \hline
        
    \end{tabular}
    \caption{\centering Subjective assessment results for contrasting pairs of districts: Thiruvarur and Thanjavur}
    \label{tab:d_analysis}
\end{table*}
\raggedbottom

Another similar example is the pair of districts of Bargarh and Sambalpur in the state of Odisha. Both are agricultural districts and had an ADI of 10 in 2011. Bargarh's ADI decreased to 9 by 2019, while Sambalpur's ADI grew rapidly to 15. Table \ref{tab:d_analysis1} shows subjective assessment results for the news articles from these districts. Articles about Sambalpur are on progressive topics such as good rainfall, job schemes, women hostels, and healthcare, while articles from Bargarh are about farmer suicides, pest attacks, and floods, among others.

\begin{table*}[]
    \centering
    \begin{tabular}{|L|L|L|L|L|L|}
         \hline
         & Agriculture & Social welfare & Environment & Industrialization & Politics and social activities\\
         \hline
        Bargarh &  
         \begin{itemize} 
            \item Crop loss
            \item Pest attack
            \item Farmer suicide due to indebtedness
        \end{itemize}
        &
        \begin{itemize}
            \item Skill institute
            \item Cancer survivor
        \end{itemize} 
        & \begin{itemize}
            \item Murders
            \item Animal deaths
            \item Flood damages
        \end{itemize} 
        & \begin{itemize}
            \item Builders fraud
            \item Mining auction
            \item Asset Raids
        \end{itemize} 
        & \begin{itemize}
            \item Political attacks
            \item Dance fest
            \item Builders fraud
        \end{itemize} 
        \\
        \hline
        Sambalpur &  
        \begin{itemize} 
            \item Good rainfall
            \item Festivals
            \item Farming activities
        \end{itemize}
        & \begin{itemize}
            \item Arrests of scam accused
            \item Rural job scheme
        \end{itemize} 
        & \begin{itemize}
            \item Tourism activities
            \item Thermal plant
            \item Save River campaign
        \end{itemize} 
        & \begin{itemize}
            \item Industri-alization effects
            \item Women hostel
            \item Develop-ment projects
        \end{itemize} 
        & \begin{itemize}
            \item Folk festivals
            \item Tourist activities
            \item Ambula-nce service
        \end{itemize} 
        \\
        \hline
        
    \end{tabular}
    \caption{\centering Subjective assessment results for contrasting pairs of districts: Bargarh and Sambalpur}
    \label{tab:d_analysis1}
\end{table*}
\raggedbottom

\subsection{Outlier analysis}
\begin{figure*}[]
    \centering
    \includegraphics[scale=0.4]{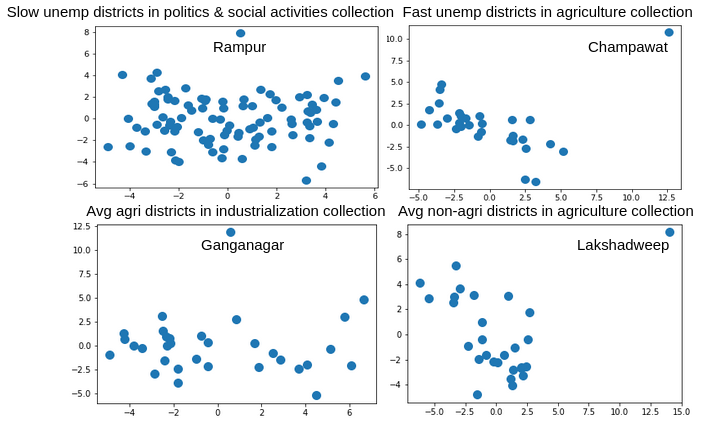}
    \caption{Outlier visualization using TSNE embeddings}
    \label{fig:tsne}
\end{figure*}

As another interesting application, we visualize using TSNE \cite{tsne} the district centroids for each sub-class and theme, and manually spot any outlier districts. The articles from these outlier districts may reveal some counter-intuitive development pathways for the sub-classes. For example, Figure \ref{fig:tsne} shows the TSNE plots for four sub-classes: based on the politics and social activities theme for slow-growing high unemployment districts, agriculture theme for fast-growing high unemployment districts, industrialization theme for fast-growing agricultural districts, and agriculture theme for medium-growing non-agricultural districts. There are some clear outliers and Table \ref{tab:out} describes the prominent topics covered in the news about these districts:


\begin{enumerate}
    \item Rampur: Upon analysing the topics covered in the news articles, this district seems to have a large number of articles about violence and murder, while the other districts in the collection prominently feature articles on politics and the development of essential amenities. 

    \item Champawat: News articles about this district discuss plans of pharmaceutical firms and government aided industrialization, while the rest of the districts see more emphasis on farming activities. Champawat therefore seems to have come up as an outlier because it is probably deviating towards an industrialization led development. 

    \item Ganganagar: Topics featured in Ganganagar are about the Information Technology sector and financial investment in agricultural industrialization, while the rest of the districts see more emphasis on core agricultural improvement such as irrigation. The outlier district therefore seems to have taken a path of attracting private sector investment for growth. 

    \item Lakshadweep: The Lakshadweep Islands are a group of islands off the Indian mainland and therefore see articles about the monsoons and cyclones, distinct from the rest of the districts. 
\end{enumerate}

This method can thus help identify some unusual districts and explain possible reasons behind their distinctive properties. 


\begin{table*}[]
    \centering
    \begin{tabular}{|M|M|M|M|M|M|}
         \hline
         Sub-class & Collection & Outlier district & Analysis on Outlier & Rest \\
         \hline
        Slow-growing unemployment districts & Politics and social activities & Rampur &
        \begin{itemize} 
            \item Mob Violence
\item Molestation
\item Murder
        \end{itemize}
&
\begin{itemize}
        \item Assembly polls
        \item Tribute to martyr
        \item Medical College
        \end{itemize} 
        \\
        \hline

         Fast-growing unemployment districts & Agriculture &  Champawat &
        \begin{itemize} 
            \item Pharma Firms
            \item Govt production plans
            \item Poll promises
        \end{itemize}
&
\begin{itemize}
            \item Flood crisis
\item Farmer protest
\item Organic farming
        \end{itemize} 
       \\
        \hline
Fast-growing agricultural districts & Industrialization & Ganganagar &
        \begin{itemize} 
            \item IT sector
            \item Upbeat investment
            \item Agro industry
        \end{itemize}
        &
        \begin{itemize}
        \itemsep0em 
            \item Cleanliness activities
\item Water lease
\item Tourist flow
        \end{itemize} 
       \\
       \hline
Average-growing non-agricultural districts & Agriculture &  Lakshadweep &
        \begin{itemize} 
            \item Cyclonic storm
            \item Pre-monsoon showers
            \item Monsoon activities
        \end{itemize}
        &
        \begin{itemize}
            \item Railway routes
\item Improve vegetable production
\item Festivals
        \end{itemize} 
       \\
        \hline
    \end{tabular}
    \caption{Subjective assessment results for outlier analysis: Rampur, Champawat, Ganganagar, Lakshadweep}
    \label{tab:out}
\end{table*}

\subsection{Unusual development pattern descriptions}

A third application we develop on similar lines as the previous application, is to identify any topics that seem unusual for a particular sub-class of development patterns. This can provide hints towards possibly new development variables that should be measured and tracked to understand their impact. We use the metric $s_{ds} - \tilde{s_{ds'}}$ to identify news articles that are representative of their sub-class but different from other sub-classes, and several surprising topics do emerge. For example, arrests of Maoist rebels and news about communal clashes is seen among slow-growing high-unemployment districts. Most districts under Maoist influence are indeed under-developed, but it is rare to find quantitative econometric research that captures the degree of Maoist activity, or issues like communal violence in many other districts, and relates them to socio-economic development \cite{maoist}. Similarly, we find that news about falling sex ratio was common among slow-growing agricultural districts, possibly indicative of growing female foeticide with poor economic development, something which again needs further investigation \cite{sexratio}. Another interesting dominant topic we identified was about cybercrime in fast-growing non-agricultural districts. This may indicate that an increasing use of digital financial services but among a newly prosperous population that is yet to become very familiar with these services, can leave the people vulnerable to financial fraud \cite{cybercrime}. Such unusual topics that emerge as dominant in a sub-class can provide hints about new variables or relationships that should be tracked to understand socio-economic development. 


%% file: newsfeed.tex
\subsection{Newsfeed for novel information}

A fourth application we demonstrate is to generate a newsfeed from among unseen (future) news articles. We consider an article about a district as providing some novel information if it deviates from typical news articles based on the development pattern of the district. Such news articles could indicate if the district is witnessing some interesting events. We do this as follows. Given a new article about a district and belonging to a particular theme, we first find its document vector using the trained DocTag2Vec model for the theme. Since the district is known to have a particular employment type (agricultural, non-agricultural, high-unemployment) and pace of growth (fast, medium, slow), we then compute the cosine similarity between the document vector of this article and the global centroid of the document vectors for the articles about that employment type. If the cosine similarity is higher between this article and an employment type different from the one known for the district, it may indicate that the district could be transitioning to a different dominant employment type. In the same manner, a similarity between this article and the global centroid for sub-classes based on the different pace of growth categories, can indicate if the district is witnessing expected events or those that might indicate a departure from the known development pattern of the district. Note that it is also possible that the media coverage about the district could be biased to cause a deviation from the socio-economic development pattern predicted through satellite based data, but we feel that the political and ideological diversity of newspapers we have included may guard against such a possibility. 

\begin{figure}
    \centering
    \includegraphics[scale = 0.25]{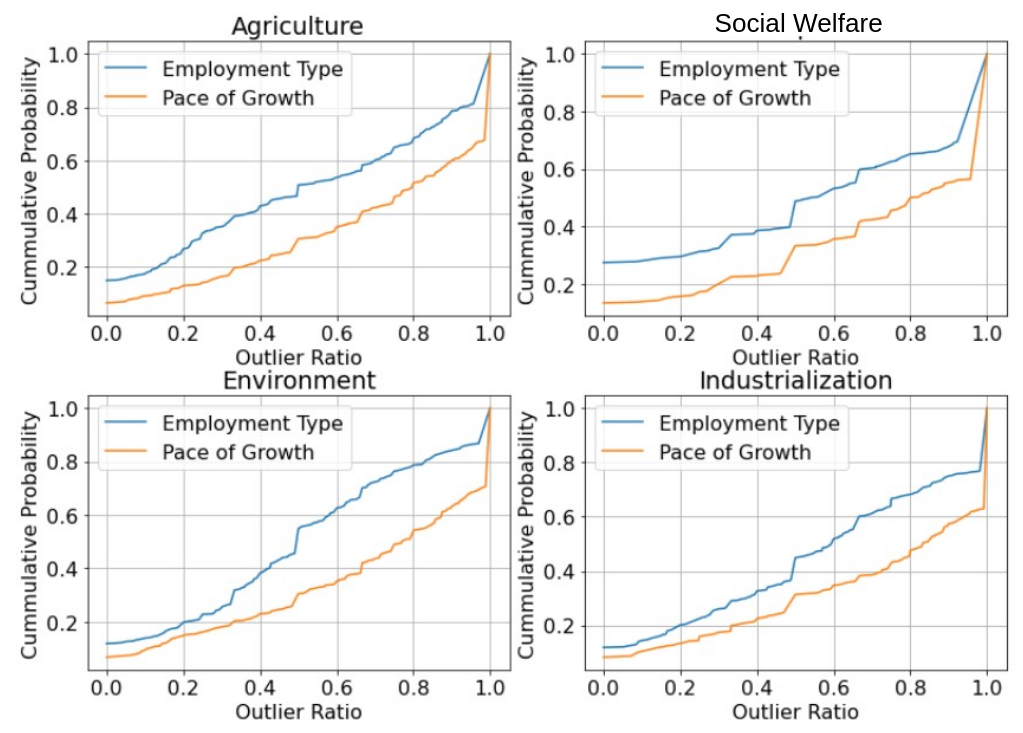}
    \caption{CDF plot for outlier ratios for Pace of Growth and Employment Type}
    \label{fig:CDF}
\end{figure}

We used the models trained on data before 2018, to generate document vectors for news articles publishing during 2018-19. Figure \ref{fig:CDF} shows CDFs of the fraction of articles in a district that appeared as outliers, i.e. the cosine similarity between the document vector of these articles was different from the document vector centroids of the employment type and the pace of growth categories to which the district belonged. Many districts indeed have a high fraction of recent news articles that diverge from the expected development pattern of the district, hinting at the need for further investigation. For example, we find that recent articles about the district of Vishakapatnam (slow-growing high unemployment district) are outliers with both the employment type outlier fraction at 0.92 and pace of growth outlier fraction at 0.87. Many of these articles are about agricultural diversification into fruits like mangoes and vegetables like bitter gourd, along with political manifestos aimed at increasing rural economic growth, possibly indicating that the district may have recently taken a turn towards improved development. In contrast, we find that articles about Jammu (fast-growing non-agricultural district) also have a high outlier fraction for the pace of growth (0.92) and are about the retraction of several aspects of Article 370 of the Indian constitution about the special autonomous status given to the state of Jammu and Kashmir. This retraction was followed with the imposition of an Internet shutdown throughout the state, and several other restrictions, likely to negatively impact the economy of the state. We thus see that such a newsfeed application can keep alerting development researchers about deviations from normal patterns, which can then be tracked and monitored more closely.

%% file: conclusion.tex
We showed in this paper that qualitative data from sources such as news articles published in the mass media can provide valuable information to explain observations made using quantitative data. We built methods using unsupervised machine learning models applied on news articles to explain different socio-economic development patterns seen by Indian districts, during an approximate ten-year span from 2011 to 2019. We demonstrated several interesting applications, such as being able to understand divergent socio-economic development patterns followed by pairs of districts that were originally similar to each other, unusual topics observed through news articles that do not conform to typical topics witnessed in a given development pattern, and a newsfeed application to aggregate and rank news articles that may be indicative of deviations of a district from development patterns it has seen so far. Our work is an effort to draw attention to the relevance of qualitative data to augment quantitative data, especially from big-data sources that build indicators for socio-economic development, and encourage similar work using other bottom-up information sources such as social media and rapid local surveys.


%% file: appendix.tex
\addcontentsline{toc}{chapter}{Appendix}
\label{ch:appendix}

\begin{enumerate}
    \item The final set of keywords to identify articles belonging to the five themes, are given in Table \ref{tab:keywords}. An article was selected for a theme only if contained at least three of these keywords. 
\begin{table}[t]
    \centering
    \begin{tabular}{|p{0.35\linewidth} | p{0.6\linewidth}|}
    \hline
         Collection & Set of keywords \\
         
         \hline
         Agriculture & agri, pesticide, insecticide, kharif crop, kharif-crop, rabi crop, rabi-crop, crops, monsoon, irrigate, farmer (includes farmer protest, farmers rally farmers
    distress), loan waiver, bhartiya kisan sangathan, bks, pradhan mantri fasal bima,
    pm fasal bima,national agriculture market, enam, pmksy, pkvy, pradhan mantri
    krishi sinchayee, pm krishi sinchayee, paramparagat krishi vikas, pm kisan
    yojna, pradhan mantri kisan yojna\\
    \hline
    Social welfare & development scheme, development program, pradhan mantri gram sadak,
    national rural employment guarantee, mgnrega, nrega, pmgsy, make in india,
    jan dhan yojna, beti bachao beti padhao, digital india, stand up india, prime
    minister ujjwala plan,pm ujjwala plan gramoday se bharat uday, shramew
    jayate , ujjwala scheme, udan, regional connectivity scheme, smart cities
    mission, skill india mission, national career service, egovernance, egov,
    aadhaar, pds, ration, nutrition, malnutrition, sanitation, hygiene, immunization,
    vaccines, ayusman bharat, rsby\\
    \hline
    
    Environment & forest, eco, environment, deforestation, wildlife, pollution, swachh bharat
    mission, swachchh bharat mission, swachhgram, clean india mission, pmfby,
    fra, integrated conservation and development, icdp, jfm, poaching, ntfp, tiger,
    leopard, zool \\
    \hline
    Industrialization & coal, lignite,steel product, industry, leather product,crude petroleum, metal
    product, textile, fertilizer, pesticide, enterprise,prime minister employment
    generation programme,estate, pmegp,credit guarantee trust fund for micro \&
    small enterprises,cgt sme,mine, mining, stock market, equity market, share
    market, factory\\
    
    \hline
    Politics and social activities & lifestyle, life-style, fashion, art, art and culture, health tips, tourism, culture,
    travel, tech, spirituality, astrology, celebrity, riot, movie, crime, violence,
    communal, hatred, fake news, misinformation, migration, suicide\\
    \hline

    \end{tabular}
    \caption{Final set of keywords used for theme-wise collections}
    \label{tab:keywords}
\end{table}

\item The parameters used for training the DocTag2Vec models were:
\begin{itemize}
        \item Learning rate: 0.1
        \item Minimum learning rate: 0.001
        \item Vector size: 50 
        \item Minimum occurrences of a word: 3
        \item Word representation: Distributed bag of words
        \item Window size for words: 3
        \item Learning rate decays linearly
\end{itemize}

\item Data from quantitative evaluation metrics for document ranking are given in Tables \ref{tab:tf_jaccard}, \ref{tab:global_cent}, \ref{tab:euclidean}, and \ref{tab:entropy}. Evaluations based on Jaccard similarity among the top-100 articles for each sub-class are shown in Table \ref{tab:tf_jaccard}, a lower value denotes more separation, and we can see that the DocTag2Vec  method gives the lowest values in most cases. Distances between articles of a sub-class and global centroids of other sub-classes, are calculated using cosine similarity in Table \ref{tab:global_cent} and Euclidean distances in Table \ref{tab:euclidean}. Lower values of cosine similarity and higher values of Euclidean distances denote greater separation. Table \ref{tab:entropy} shows results from an entropy-based measure, where the entropy of a sub-class is calculated based on the probability of keywords belonging more to that sub-class than other competing sub-classes. Methods producing lower values of entropy are better.



\begin{table*}[]
    \centering
    \begin{tabular}{|c|c|c|c|c|c|L|}
    \hline
                    &   & Agriculture & Social welfare & Environment & Industrialization & Politics and social activities\\
    \hline  
    \multirow{3}{*}{Slow vs fast unemp} & DocTag2Vec Method &    0.12 & 0.13 & 0.13 & 0.12 & 0.11 \\
    \cline{2-7}
                                 & \begin{minipage}{1.2in} \strut \centering TFIDF based selection  \strut \end{minipage} &     0.12 & 0.13 & 0.14 & 0.14 & 0.12\\
    \cline{2-7}
                    & LDA           &     0.14 & 0.12 & 0.13 & 0.14 & 0.12\\
    \hline
    \multirow{3}{*}{Slow vs fast agri} & DocTag2Vec Method &    0.14 & 0.15 & 0.12 & 0.11 & 0.14\\
    \cline{2-7}
                                 & \begin{minipage}{1.2in} \strut \centering TFIDF based selection  \strut \end{minipage} & 0.15 & 0.14 & 0.12 & 0.13 & 0.12 \\
    
    \cline{2-7}
                         & LDA           &      0.15 & 0.15 & 0.13 & 0.11 & 0.12\\
    \hline
    \multirow{3}{*}{Slow vs fast non-agri} & DocTag2Vec Method & 0.13 & 0.13 & 0.13 & 0.13 & 0.11 \\
    \cline{2-7}
                                 & \begin{minipage}{1.2in} \strut \centering TFIDF based selection \strut \end{minipage} & 0.12 & 0.13 & 0.12 & 0.12 & 0.13\\
    
    \cline{2-7}
                           & LDA           &    0.12 & 0.13 & 0.14 & 0.09 & 0.12\\
    \hline

    \end{tabular}
    \caption{Jaccard similarity metric}
    \label{tab:tf_jaccard}
\end{table*}
\raggedbottom




\begin{table*}[]
    \centering
    \begin{tabular}{|c|c|c|c|c|c|L|}
    \hline
                    &   & Agriculture & Social welfare & Environment & Industrialization & Politics and social activities\\
    \hline  
    \multirow{3}{*}{Slow vs fast unemp} & DocTag2Vec Method & 0.66 & 0.71 & 0.67 & 0.76 & 0.61 \\
    \cline{2-7}
                                 & \begin{minipage}{1.2in} \strut \centering TFIDF based selection  \strut \end{minipage} & 0.77 & 0.61 & 0.76 & 0.8 & 0.7\\
    \cline{2-7}
                                 & LDA           & 0.78 & 0.66 & 0.78 & 0.76 & 0.71\\
    
    \hline
    \multirow{3}{*}{Slow vs fast agri} & DocTag2Vec Method & 0.64 & 0.58 & 0.66 & 0.73 & 0.65 \\ 
    \cline{2-7}
                                 & \begin{minipage}{1.2in} \strut \centering TFIDF based selection  \strut \end{minipage} &   0.67 & 0.72 & 0.65 & 0.77 & 0.74\\
    \cline{2-7}
                                 & LDA           &  0.63 & 0.71 & 0.70 & 0.80 & 0.72 \\
    
    \hline
    \multirow{3}{*}{Slow vs fast non-agri} & DocTag2Vec Method & 0.44 & 0.43 & 0.29 & 0.51 & 0.44  \\
    \cline{2-7}
                                 & \begin{minipage}{1.2in} \strut \centering TFIDF based selection \strut \end{minipage} &  0.37 & 0.37 & 0.44 & 0.56 & 0.47 \\
    \cline{2-7}
                                 & LDA           &  0.51 & 0.40 & 0.56 & 0.45 & 0.41 \\
    
    \hline

    \end{tabular}
    \caption{Global centroid similarity metric}
    \label{tab:global_cent}
\end{table*}
\raggedbottom



\begin{table*}[]
    \centering
    \begin{tabular}{|c|c|c|c|c|c|L|}
    \hline
                    &   & Agriculture & Social welfare & Environment & Industrialization & Politics and social activities\\
    \hline  
    \multirow{3}{*}{Slow vs fast unemp} & DocTag2Vec Method &  4.33 & 4.24 & 4.80 & 4.86 & 5.05\\
    \cline{2-7}
                                 & \begin{minipage}{1.2in} \strut \centering TFIDF based selection  \strut \end{minipage} &    3.9 & 5.43 & 4.67 & 3.98 & 4.85 \\
    \cline{2-7}
                                 & LDA           &  3.49 & 4.29 & 3.85 & 4.30 & 4.38\\
    
    \hline
    \multirow{3}{*}{Slow vs fast agri} & DocTag2Vec Method &   4.88 & 4.55 & 4.41 & 5.16 & 5.46 \\
    \cline{2-7}
                                 & \begin{minipage}{1.2in} \strut \centering TFIDF based selection  \strut \end{minipage} &    5.4 & 4.09 & 4.91 & 4.8 & 4.2\\
    \cline{2-7}
                                 & LDA           &    4.74 & 4.23 & 4.72 & 4.99 & 5.49\\
    
    \hline
    \multirow{3}{*}{Slow vs fast non-agri} & DocTag2Vec Method &   5.04 & 5.33 & 6.53 & 5.20 & 4.34 \\
    \cline{2-7}
                                 & \begin{minipage}{1.2in} \strut \centering TFIDF based selection \strut \end{minipage} &  5.66 & 5.63 & 4.71 & 5.3 & 5.47 \\
    \cline{2-7}
                                 & LDA           & 4.94 & 4.95 & 3.92 & 4.96 & 4.85\\
    
    \hline

    \end{tabular}
    \caption{Euclidean distance metric}
    \label{tab:euclidean}
\end{table*}
\raggedbottom



\begin{table*}[]
    \centering
    \begin{tabular}{|c|c|c|c|c|c|L|}
    \hline
                    &   & Agriculture & Social welfare & Environment & Industrialization & Politics and social activities\\
    \hline  
    \multirow{3}{*}{Slow vs fast unemp} & DocTag2Vec Method & 5.96 & 5.96 & 5.84 & 5.73 & 6.0 \\
    \cline{2-7}
                                 & \begin{minipage}{1.2in} \strut \centering TFIDF based selection  \strut \end{minipage} & 5.57 & 6.08 & 5.77 & 5.94 & 5.69\\
    \cline{2-7}
                                 & LDA           & 6.02 & 6.03 & 5.89 & 5.92 & 5.95\\
    
    \hline
    \multirow{3}{*}{Slow vs fast agri} & DocTag2Vec Method & 5.77 & 6.06 & 5.94 & 5.97 & 5.73 \\ 
    \cline{2-7}
                                 & \begin{minipage}{1.2in} \strut \centering TFIDF based selection  \strut \end{minipage} &   5.80 & 5.78 & 5.84 & 4.73 & 5.96\\
    \cline{2-7}
                                 & LDA           &  5.78 & 5.93 & 5.84 & 6.01 & 5.88 \\
    
    \hline
    \multirow{3}{*}{Slow vs fast non-agri} & DocTag2Vec Method & 5.86 & 6.01 & 5.95 & 5.98 & 5.95  \\
    \cline{2-7}
                                 & \begin{minipage}{1.2in} \strut \centering TFIDF based selection \strut \end{minipage} & 5.91 & 5.98 & 5.99 & 5.87 & 5.92\\
    \cline{2-7}
                                 & LDA           &  5.96 & 5.96 & 5.74 & 5.82 & 6.0 \\
    
    \hline

    \end{tabular}
    \caption{Entropy metric}
    \label{tab:entropy}
\end{table*}
\raggedbottom


\item Qualitative evaluation: Standard two sample t-tests were performed on the review ratings given by eleven raters to the document selections obtained from the three different methods. A t-test was done to compare these methods in pairs with each other, and the tests were done separately for each question out of the three questions put up to the raters. The null hypothesis that pairs of ratings were similar to each other, was rejected in most cases with p-values less than 0.05. Table \ref{tab:t_test} shows the results. The questions put up to the raters are shown in Table \ref{tab:review}. 


\begin{table*}[]
    \centering
    \tiny
    \begin{tabular}{|N|N|N|N|N|N|N|N|N|N|N|N|N|N|N|N|N|N|N|N|N|N|}
        \hline
         & & \multicolumn{4}{c|}{Agriculture} & \multicolumn{4}{c|}{Development} & \multicolumn{4}{c|}{Environment} & \multicolumn{4}{c|}{Industrialization} & \multicolumn{4}{c|}{Politics and social activities} \\
        \hline
        & & \multicolumn{2}{c|}{p1=DocTag2Vec} & \multicolumn{2}{c|}{p1=DocTag2Vec} & \multicolumn{2}{c|}{p1=DocTag2Vec} & \multicolumn{2}{c|}{p1=DocTag2Vec}  & \multicolumn{2}{c|}{p1=DocTag2Vec} & \multicolumn{2}{c|}{p1=DocTag2Vec}    & \multicolumn{2}{c|}{p1=DocTag2Vec} & \multicolumn{2}{c|}{p1=DocTag2Vec} & \multicolumn{2}{c|}{p1=DocTag2Vec} & \multicolumn{2}{c|}{p1=DocTag2Vec} \\
         & & \multicolumn{2}{c|}{p2=TFIDF} & \multicolumn{2}{c|}{p2=LDA} & \multicolumn{2}{c|}{p2=TFIDF} & \multicolumn{2}{c|}{p2=LDA}  & \multicolumn{2}{c|}{p2=TFIDF} & \multicolumn{2}{c|}{p2=LDA}    & \multicolumn{2}{c|}{p2=TFIDF} & \multicolumn{2}{c|}{p2=LDA} & \multicolumn{2}{c|}{p2=TFIDF} & \multicolumn{2}{c|}{p2=LDA} \\
    \hline
    
	& & t-value & p-value & t-value & p-value & t-value & p-value & t-value & p-value & t-value & p-value & t-value & p-value & t-value & p-value & t-value & p-value & t-value & p-value & t-value & p-value \\
          
    \hline
    
	\multirow{3}{0.5cm}{Slow vs Fast unemp} &  Q1 & 1.112 & 0.279 & 0.933 & 0.362 & 2.196 & 0.04 & 1.077 & 0.294 & 1.627 & 0.119 & 1.978 & 0.062 & 4.041 & 0.001 & 6.264 & 0.0004 & 3.149 & 0.005 & 2.864 & 0.01 \\

    \cline{2-22}

	& Q2 & 2.149 & 0.044 & 2.344 & 0.03 & 0.806 & 0.429 & 0.0 & 1.0 & 1.102 & 0.284 & 0.464 & 0.648 & 2.229 & 0.037 & 2.752 & 0.012 & 2.569 & 0.018 & 2.372 & 0.028 \\
    \cline{2-22}

	& Q3 & 2.401 & 0.026 & 3.305 & 0.004 & -0.501 & 0.622 & 0.264 & 0.794 & 0.761 & 0.455 & 1.204 & 0.243 & 1.668 & 0.111 & 1.0 & 0.329 & 1.479 & 0.155 & 1.352 & 0.191 \\

    \hline
	\multirow{3}{0.5cm}{Slow vs Fast agri} & Q1 & 3.796 & 0.001 & 3.551 & 0.002 & -1.373 & 0.185 & 0.753 & 0.46 & 4.008 & 0.001 & 2.911 & 0.009 & 2.142 & 0.045 & 2.397 & 0.026 & 0.267 & 0.792 & -1.432 & 0.168 \\

    \cline{2-22}
	& Q2 &  2.548 & 0.019 & 4.561 & 0.0 & -0.801 & 0.433 & 0.598 & 0.557 & 3.227 & 0.004 & 2.813 & 0.011 & 1.593 & 0.127 & 0.767 & 0.452 & 0.755 & 0.459 & -0.503 & 0.621 \\

    \cline{2-22}

	& Q3 & 3.212 & 0.004 & 1.0 & 0.329 & 1.708 & 0.103 & 1.006 & 0.326 & 1.48 & 0.154 & 1.725 & 0.1 & 2.677 & 0.014 & 2.236 & 0.037 & 1.204 & 0.243 & 0.977 & 0.34 \\
    \hline
	\multirow{3}{0.7cm}{Slow vs Fast non-agri} & Q1 &  3.493 & 0.002 & 4.472 & 0.0002 & 2.41 & 0.026 & 1.021 & 0.32 & 3.117 & 0.005 & 0.75 & 0.462 & 2.85 & 0.01 & 3.07 & 0.006 & 6.708 & 0.0001 & 3.833 & 0.001 \\
    \cline{2-22}

	& Q2 & 2.57 & 0.018 & 3.774 & 0.001 & 2.381 & 0.027 & 0.169 & 0.867 & 2.193 & 0.04 & 1.472 & 0.156 & 3.608 & 0.002 & 4.146 & 0.0001 & 2.008 & 0.058 & 1.48 & 0.154 \\

    \cline{2-22}
	& Q3 & 0.342 & 0.736 & 0.775 & 0.447 & -0.439 & 0.665 & 0.706 & 0.489 & -0.644 & 0.527 & 1.337 & 0.196 & 0.0 & 1.0 & -0.338 & 0.739 & 1.513 & 0.146 & 2.665 & 0.015 \\

	\hline
          
    \end{tabular}
    \caption{T-test results from qualitative evaluation}
    \label{tab:t_test}
\end{table*}

\item A subjective assessment of the patterns is given in Tables \ref{tab: pat1}, \ref{tab: pat2}, and \ref{tab: pat3}, for the three patterns, done by manually examining the dominant topics covered in the top-100 articles for each sub-class. 

\begin{table*}[t]
    \centering
    \tiny
    \begin{tabular}{|L|L|L|L|L|L|L|}
         \hline
          & & Agriculture & Social welfare & Environment & Industrialization & Politics and social activities\\
         \hline
          DocTag2Vec method & Slow growing Unemployment districts & 
         \begin{itemize}
             \item Food godowns
             \item Policy execution
             \item Heavy rainfall
         \end{itemize}
         &
         \begin{itemize}
             \item Irregularities in job scheme
             \item Compensation to killed
             \item Policy implementation reviews
         \end{itemize}  
         &
         \begin{itemize}
             \item Pitch for quality education
             \item Springs under threat
             \item Review power pyre
         \end{itemize}
         &
         \begin{itemize}
             \item Gang wars, Thefts
            \item Free wifi
            \item Growth problems
         \end{itemize}
         &
         \begin{itemize}
             \item Trade Fair
            \item Healthcare promise
            \item Jail Raid 
         \end{itemize}\\
         \cline{2-7}
        &  Fast growing Unemployment districts & 
         \begin{itemize}
             \item Production improvement 
            \item Farmer Suicide
            \item Rain \& Floods
         \end{itemize}
         &
         \begin{itemize}
             \item Health Fair
            \item Development projects
            \item Industrial progress

         \end{itemize}  
         &
         \begin{itemize}
             \item Industrial fuel
            \item Wildlife protection
            \item Air purifiers

         \end{itemize}
         &
         \begin{itemize}
             \item Biomining project
            \item Fine for burning
            \item Air pollution

         \end{itemize}
         &
         \begin{itemize}
             \item Curb crime
            \item Textile celebrations
            \item Farming experience

         \end{itemize}\\
         \hline
         
          TFIDF based selection & Slow growing Unemployment districts & 
         \begin{itemize}
            \item Political events
\item Power plants
\item Dalit issues

         \end{itemize}
         &
         \begin{itemize}
             \item Crimes against women
\item Digital India
\item Financial assistance
         \end{itemize}  
         &
         \begin{itemize}
        \item Free wifi
\item Toilets
\item Environment Day    
        \end{itemize}
         &
         \begin{itemize}
        \item Holy Festival
\item Illegal mining
\item Ailing industry
         \end{itemize}
         &
         \begin{itemize}
        \item Riots
\item Deaths
\item Monument protection
         \end{itemize}\\
         \cline{2-7}
          & Fast growing Unemployment districts & 
         \begin{itemize}
        \item Floods
\item Health risks
\item Pesticides
         \end{itemize}
         &
         \begin{itemize}
        \item MGNREGA
\item Investment activities
\item Education problems
         \end{itemize}  
         &
         \begin{itemize}
         \item Air Pollution
\item Hunter activities
\item Clean Ganga
         \end{itemize}
         &
         \begin{itemize}
           \item Technology
\item Cyclone recovery
\item Industrial growth
         \end{itemize}
         &
         \begin{itemize}
           \item Bad air quality
\item Murders
\item Democracy crisis

         \end{itemize}\\
         \hline
          LDA & Slow growing Unemployment districts & 
         \begin{itemize}
             \item Agri production
\item Development packages
\item Political march
         \end{itemize}
         &
         \begin{itemize}
             \item Air connectivity
\item Plantations
\item Rural development
         \end{itemize}  
         &
         \begin{itemize}
            \item Highway alignment
\item Development project
\item Tiger Reserve
         \end{itemize}
         &
         \begin{itemize}
            \item Agitation
\item Farming
\item Chit fund scam
         \end{itemize}
         &
         \begin{itemize}
           \item Health camp
\item Unlawful activities
\item Riots
         \end{itemize}\\
         \cline{2-7}
        &  Fast growing Unemployment districts & 
         \begin{itemize}
            \item Farmer death
\item Seed subsidy
\item Awards
         \end{itemize}
         &
         \begin{itemize}
             \item Agri business
\item MGNREGA
\item Mass marriage

         \end{itemize}  
         &
         \begin{itemize}
            \item NGO activities
\item Suicide
\item Zoo animals
         \end{itemize}
         &
         \begin{itemize}
            \item Cleanliness drive
\item Property stealing
\item Deaths

         \end{itemize}
         &
         \begin{itemize}
             \item Riots
\item Infra development plan
\item Plastic ban

         \end{itemize}\\
         \hline

    \end{tabular}
    \caption{Subjective assessment results for Pattern 1}
    \label{tab: pat1}
\end{table*}

\begin{table*}[t]
    \centering
    \tiny
    \begin{tabular}{|L|L|L|L|L|L|L|}
         \hline
          & & Agriculture & Social welfare & Environment & Industrialization & Politics and social activities\\
         \hline
          DocTag2Vec method & Slow growing Agricultural districts & 
         \begin{itemize}
             \item Rain \& Floods
\item Farmer deaths
\item Arts expo

         \end{itemize}
         &
         \begin{itemize}
            \item Farmer politics
\item Pending wage payments
\item Raise in air fare 
         \end{itemize}  
         &
         \begin{itemize}
            \item Fuel adulteration
\item Rapes
\item Pollution
         \end{itemize}
         &
         \begin{itemize}
         \item Power staff strike
\item Bus service suspended
\item Sanitation drive

         \end{itemize}
         &
         \begin{itemize}
        \item Protests
\item Violence \& Crimes
\item Art \& Design

         \end{itemize}\\
         \cline{2-7}
         & Fast growing Agricultural districts & 
         \begin{itemize}
             \item Crop loss compensation
\item Hike in power tariff
\item Best practices for farmers

         \end{itemize}
         &
         \begin{itemize}
\item Conservation projects
\item Plantation scheme
\item Intellectual property meet

         \end{itemize}  
         &
         \begin{itemize}
             \item Protest against poaching
\item River protection petition
\item Wild-life habitat

         \end{itemize}
         &
         \begin{itemize}
            \item Mining activities
\item Milk price hike
\item Transportation activites

         \end{itemize}
         &
         \begin{itemize}
                \item Tech Summit
    \item Celebrations
    \item Elections

         \end{itemize}\\
         \hline
         
          TFIDF based selection & Slow growing Agricultural districts & 
         \begin{itemize}
            \item Food prices
\item Water conservation
\item Better harvest

         \end{itemize}
         &
         \begin{itemize}
          \item Job scheme
\item Boost agriculture
\item MGNREGA dilution
         \end{itemize}  
         &
         \begin{itemize}
        \item Forest degradation
\item Tiger reserve
\item Waste water  
        \end{itemize}
         &
         \begin{itemize}
        \item Violence
\item Smart apps
\item Coal mine
         \end{itemize}
         &
         \begin{itemize}
        \item Train services
\item Cow violence
\item Liquor
         \end{itemize}\\
         \cline{2-7}
          & Fast growing Agricultural districts & 
         \begin{itemize}
        \item Water industry
\item Fodder substitute
\item Drought
         \end{itemize}
         &
         \begin{itemize}
        \item Aviation policy
\item Business expo
\item Air connectivity
         \end{itemize}  
         &
         \begin{itemize}
         \item Infra project
\item Awareness on afforestation
\item Fight against pollution
         \end{itemize}
         &
         \begin{itemize}
          \item Income Tax raids
\item Water crisis
\item Mining activities
         \end{itemize}
         &
         \begin{itemize}
          \item Narcotics
\item Protests
\item Bullet train

         \end{itemize}\\
         \hline
         LDA & Slow growing Agricultural districts & 
         \begin{itemize}
            \item Monsoon
\item Farm crisis
\item Price rise

         \end{itemize}
         &
         \begin{itemize}
            \item Village development
\item Skill programme
\item Water project
         \end{itemize}  
         &
         \begin{itemize}
            \item Tourism
\item Killings
\item Reservoirs overflow
         \end{itemize}
         &
         \begin{itemize}
         \item Free wifi
\item Illegal mining
\item Protest march

         \end{itemize}
         &
         \begin{itemize}
        \item Murder
\item Suspension
\item Entrance exam

         \end{itemize}\\
         \cline{2-7}
         & Fast growing Agricultural districts & 
         \begin{itemize}
             \item Hill highway
\item MGNREGA
\item Drought

         \end{itemize}
         &
         \begin{itemize}
\item Rail budget
\item Development MoUs
\item Taxation

         \end{itemize}  
         &
         \begin{itemize}
             \item Tourist spot
\item Road project
\item Textile activities

         \end{itemize}
         &
         \begin{itemize}
           \item Waste segregation
\item Telecom firms
\item Blast suspect
         \end{itemize}
         &
         \begin{itemize}
              \item Road accident
\item Health check-up
\item Arrests

         \end{itemize}\\
         \hline

    \end{tabular}
    \caption{Subjective assessment results for Pattern 2}
    \label{tab: pat2}
\end{table*}

\begin{table*}[t]
    \centering
    \tiny
    \begin{tabular}{|L|L|L|L|L|L|L|}
         \hline
          & & Agriculture & Social welfare & Environment & Industrialization & Politics and social activities\\
         \hline
          DocTag2Vec method & Slow growing Non-Agricultural districts & 
         \begin{itemize}
             \item Rain \& Floods
\item Farmer deaths
\item Arts expo

         \end{itemize}
         &
         \begin{itemize}
          \item Folk programmes
\item Awareness training
\item Pension scheme

         \end{itemize}  
         &
         \begin{itemize}
            \item Plantation drive
\item Poor air quality
\item Clean Ganga cursade
         \end{itemize}
         &
         \begin{itemize}
         \item Demonetization
\item Power crisis
\item Illegal mining

         \end{itemize}
         &
         \begin{itemize}
        \item Elections
\item Violence
\item Politics
         \end{itemize}\\
         \cline{2-7}
        &  Fast growing Non-Agricultural districts & 
         \begin{itemize}
            \item Farmers immolation
\item Capital city project
\item Strike warning

         \end{itemize}
         &
         \begin{itemize}
\item Air link scheme
\item Anti-tank missiles
\item Firm activities

         \end{itemize}  
         &
         \begin{itemize}
             \item Production plant
\item Pollution spike
\item Wildlife tourists

         \end{itemize}
         &
         \begin{itemize}
            \item Economy revival
\item Infra growth
\item Revenue generation

         \end{itemize}
         &
         \begin{itemize}
              \item Modern economy threat
\item Hacking
\item Politics

         \end{itemize}\\
         \hline
         
          TFIDF based selection & Slow growing Non-Agricultural districts & 
         \begin{itemize}
            \item Air quality
\item Demonetisation
\item Elections
         \end{itemize}
         &
         \begin{itemize}
         \item Connectivity scheme
\item Debt waiver scheme
\item Transportation
         \end{itemize}  
         &
         \begin{itemize}
        \item Child rights
\item Air pollution
\item Urban planning 
        \end{itemize}
         &
         \begin{itemize}
        \item GST impact
\item CEAT company
\item Co-living spaces
         \end{itemize}
         &
         \begin{itemize}
        \item Accidential deaths
\item Protest activities
\item Politics
         \end{itemize}\\
         \cline{2-7}
          & Fast growing Non-Agricultural districts & 
         \begin{itemize}
        \item Dam Safety Bill
\item Pollution
\item Rainfalls
         \end{itemize}
         &
         \begin{itemize}
        \item Food prices
\item Policy change
\item National meet

         \end{itemize}  
         &
         \begin{itemize}
         \item Protest site
\item Influence of rivers
\item Politics
         \end{itemize}
         &
         \begin{itemize}
          \item PM speeches
\item Criminal activities
\item Banking sector

         \end{itemize}
         &
         \begin{itemize}
         \item Cricket
\item Election results
\item Underworld crimes

         \end{itemize}\\
         \hline
         LDA & Slow growing Non-Agricultural districts & 
         \begin{itemize}
             \item River link
\item Medical University
\item Drains

         \end{itemize}
         &
         \begin{itemize}
          \item E-toilets
\item Consulting firms
\item Housing scheme
         \end{itemize}  
         &
         \begin{itemize}
            \item Pollution
\item Green plan
\item LPG scheme
         \end{itemize}
         &
         \begin{itemize}
         \item Illegal mining
\item Telecom providers
\item Insurance firms

         \end{itemize}
         &
         \begin{itemize}
        \item Disputes
\item Rain
\item Arrests
         \end{itemize}\\
         \cline{2-7}
        &  Fast growing Non-Agricultural districts & 
         \begin{itemize}
            \item Suicide
\item Drought
\item Food bill

         \end{itemize}
         &
         \begin{itemize}
\item Thermal unit
\item Fund release
\item Tourism

         \end{itemize}  
         &
         \begin{itemize}
            \item Mining
\item Youth Fest
\item Debt markets
         \end{itemize}
         &
         \begin{itemize}
            \item Telcos
\item Drug prices
\item Bank norms

         \end{itemize}
         &
         \begin{itemize}
              \item Crime
\item Health programme
\item Sensex

         \end{itemize}\\
         \hline
    \end{tabular}
    \caption{Subjective assessment results for Pattern 3}
    \label{tab: pat3}
\end{table*}


\begin{table*}
    \centering
    \begin{tabular}{|c|c|}
    
         \hline
         Collection & Questions \\
         \hline
         Agriculture & 
         \begin{minipage}{12cm}
         \begin{itemize}
             \item Are the articles depicting pace of growth i.e. slow vs fast in agriculture sector like slow pace is describing about floods, monsoon, droughts, suicides while fast is describing about compensations, crop prices, relief schemes etc? 
\item Are the articles relevant to the sub-classes i.e. are they discussing about farming activities, crops, compensations to farmers etc?
\item Are the topics discussed in sub-classes different from each other?
\end{itemize}
         \end{minipage}

         \\
         \hline
          Social welfare & 
           \begin{minipage}{12cm}
         \begin{itemize}
             \item Are the articles depicting pace of growth i.e. slow vs fast in terms of development as slow ones are talking about failure in development schemes, unemployment and fast is talking about success of schemes and new schemes? 
\item Are the articles relevant to the sub-classes i.e. they are talking about job schemes, investments, development activities?
\item Are the topics discussed in sub-classes different from each other?
         \end{itemize}
            \end{minipage} 
         
         \\
         \hline
         Environment & 
          \begin{minipage}{12cm}
         \begin{itemize}
             \item Are the articles depicting pace of growth i.e. slow vs fast in terms of environment as slow ones are talking about environmental problems as pollution, animal killing while fast is talking about activities for wildlife protection, prevention of poaching activities? 
\item Are the articles relevant to the sub-classes i.e. slow ones are talking about environmental and wildlife problems while fast ones are talking about steps towards their protection?
\item Are the topics discussed in sub-classes different from each other?
        \end{itemize}
         
         \end{minipage}
         \\
         
         \hline
         Industrialization & 
          \begin{minipage}{12cm}
         \begin{itemize}
             \item Are the articles depicting pace of growth i.e. slow vs fast in terms of industrialization as slow ones are talking about problems in industry, failed investments while fast ones are focusing on development activities in industry as revenue generation, trades etc? 
\item Are the articles relevant to the sub-classes i.e. slow ones are focusing on problems related to different industrial sectors while fast ones are inclined towards industrial development, investment, schemes?
\item Are the topics discussed in sub-classes different from each other?
\end{itemize}
         \end{minipage}

         \\
         \hline
        Politics and social activities & 
         \begin{minipage}{12cm}
         \begin{itemize}
             \item Are the articles depicting pace of growth i.e. slow vs fast in terms of lifestyle like slow ones are focusing on women problems, violence, crimes while fast ones are focusing on celebrations, fight against crimes, special cells ? 
\item Are the articles relevant to the sub-classes i.e. slow is talking about crimes, rapes, arrests while fast is talking about police activities against crime, law and order, celebrations etc?
\item Are the topics discussed in sub-classes different from each other?

        \end{itemize}
         
         \end{minipage}
         \\
         
         \hline
    \end{tabular}
    \caption{Questionnaire for qualitative review}
    \label{tab:review}
\end{table*}

\end{enumerate}